ORIGINAL ARTICLE

# DP-SMOTe: Integrating Differential Privacy and Oversampling Technique to Preserve Privacy in Smart Homes

Amr Tarek Elsayed [a,*], Almohammady Sobhi Alsharkawy [a], Mohamed Sayed Farag [b], Shaban Ebrahim Abu Yusuf [c]

[a] Computer Science Department, Faculty of Science (Boys), Al-Azhar University, Cairo, Egypt
[b] Computer Science Department, Obour High Institute for Informatics, Cairo, Egypt
[c] Department of Mathematics, Faculty of Science (Boys), Al-Azhar University, Cairo, Egypt

## Abstract

Smart homes represent intelligent environments where interconnected devices gather information, enhancing users' living experiences by ensuring comfort, safety, and efficient energy management. To enhance the quality of life, companies in the smart device industry collect user data, including activities, preferences, and power consumption. However, sharing such data necessitates privacy-preserving practices. This paper introduces a robust method for secure sharing of data to service providers, grounded in differential privacy (DP). This empowers smart home residents to contribute usage statistics while safeguarding their privacy. The approach incorporates the Synthetic Minority Oversampling technique (SMOTe) and seamlessly integrates Gaussian noise to generate synthetic data, enabling data and statistics sharing while preserving individual privacy. The proposed method employs the SMOTe algorithm and applies Gaussian noise to generate data. Subsequently, it employs a k-anonymity function to assess re-identification risk before sharing the data. The simulation outcomes demonstrate that our method delivers strong performance in safeguarding privacy and in accuracy, recall, and f-measure metrics. This approach is particularly effective in smart homes, offering substantial utility in privacy at a re-identification risk of 30%, with Gaussian noise set to 0.3, SMOTe at 500%, and the application of a k-anonymity function with k = 2. Additionally, it shows a high classification accuracy, ranging from 90% to 98%, across various classification techniques.

## 1. Introduction

The increasing appeal of smart residences reflects a growing desire among individuals to have greater control over their living spaces and enhance their overall quality of life. Within a smart home, an array of devices, including but not limited to smart thermostats, security systems, lighting, and entertainment systems, can be seamlessly operated from afar. The connectivity of these devices to the internet enables users to remotely administer them using their cellphones or other internet-connected devices [1]. Perhaps the most noteworthy aspect of a smart home is its remarkable convenience. It grants users the power to govern the temperature, lights, and security of their dwelling from virtually any location worldwide. Leveraging a smartphone, one can effortlessly extinguish lights, adjust the thermostat, and monitor security cameras, ensuring both the comfort and safety of the home. Additionally, smart homes deliver tangible savings in terms of energy costs. Smart thermostats, for instance, are adept at adjusting a home's temperature in alignment with the inhabitants' preferences and presence, thus translating into reduced expenditures on heating and cooling. Similarly, smart lighting systems possess the capability to automatically power down lights when rooms remain unoccupied, leading to significant energy conservation [2].

Smart homes offer enhanced security features, allowing homeowners to oversee their properties from remote locations and receive alerts in case of suspicious activities, thus augmenting protection [3]. Furthermore, the capacity to lock and unlock doors from a distance permits the convenient admission of guests or service personnel without physical presence.

In the realm of entertainment, smart homes have the ability to elevate one's leisure experiences. Control over television, music, and other entertainment systems can be exerted from any point within the smart home. The integration of voice assistants further simplifies entertainment management through voice commands. The wealth of data harvested from smart homes holds significant potential for a wide array of applications, spanning from the prediction of smart home activities [4] to improved healthcare services for patient treatment [5], enabling more effective disorder assessment, enhancing smart city pedestrian monitoring systems [6], and optimizing energy management strategies. This data is increasingly recognized by businesses as a valuable resource for enhancing their products and services.

However, it is imperative for data collectors to prioritise the confidentiality of such data. Mishandling or inadequate management of this data can give rise to substantial issues. As a result, a novel system has been proposed, one that harmonizes both privacy and utility. In the realm of remote health systems, the processes of collecting, disclosing, and utilizing personal health information give rise to significant privacy concerns. Households are often perceived as the utmost private environments by many individuals. In these settings, devices like glucometers for blood sugar measurement, spirometers for lung function assessment, and sensors monitoring sleep patterns are commonly used. These devices can inadvertently disclose sensitive health information, potentially indicating conditions such as diabetes, asthma, or depressive disorders. Consequently, patients frequently prefer to limit access to this data, usually confining it to a selected group, predominantly their personal healthcare providers, due to privacy considerations.

Differential Privacy, as highlighted in [7], has emerged as a widely embraced technique for safeguarding privacy. Its core concept involves granting users a level of plausible deniability by introducing random values into their input. This methodology, in the realm of centralized differential privacy, fortifies user privacy significantly, thereby shielding their data from potential adversaries, including service providers and external parties. In this scenario, a process entails the introduction of noise into the database, complemented by the application of a differential privacy aggregation algorithm.

The Synthetic Minority Over-sampling Technique (SMOTe), introduced in [8], is employed to address imbalanced datasets by creating synthetic samples for the minority class, thereby mitigating bias in machine learning models towards the majority class. When coupled with privacy-preserving methodologies, SMOTe plays a pivotal role in fortifying the privacy of minority class data. By generating synthetic instances while upholding data confidentiality, it contributes to the development of more equitable and secure predictive models.

This paper presents an innovative approach to securely transmit household data to the aggregator while addressing potential threats posed by malicious aggregator nodes. To tackle this issue, we employ Differential Privacy to safeguard real-time data collected from residences. Before transmission, a privacy preservation process is applied, utilizing the SMOTe algorithm to generate synthetic data and adding Gaussian noise to the generated data. This guarantees that the aggregator is unable to determine the identity of the individual resident, thereby maintaining their anonymity. The proposed model presents unique benefits over current approaches. By employing SMOTe and Gaussian noise, it creates a protected and anonymous data setting, successfully preventing any hostile aggregator nodes from obtaining confidential data.

The main contributions of this paper are as follows:

- Presenting a well-defined formulation specifically tailored to address the challenges of privacy preservation in smart homes.
- Employing the Synthetic Minority Over-sampling Technique (SMOTe) algorithm to generate synthetic datasets.
- Enhancing data privacy by applying Gaussian noise to the dataset.
- Implementing various classification methods, to ascertain that the accuracy of the data is not substantially compromised post-application of our privacy-preserving techniques. This procedure validates the utility of the generated data, confirming that our approach effectively balances privacy with the need for accurate data analysis.
- Utilizing the k-anonymity model as a pivotal framework for conducting security analysis.

The subsequent sections of this paper are organized as follows: Section 2 reviews current privacy preservation techniques in smart homes and underscores their limitations. Section 3 provides essential background information on SMOTe and Gaussian noise. Section 4 outlines our approach and introduces the system model. In Section 5, the scheme's performance is analyzed from both security and efficiency perspectives.

## 2. Related work

In recent times, the intersection of data synthesis methods with Differential Privacy (DP) solutions has been a pivotal area of research. This convergence aims to address the challenge of releasing data for analysis, ensuring its usefulness while simultaneously safeguarding the privacy of individuals [9,10]. Generative Adversarial Networks (GAN) are widely recognized for their capacity to produce synthetic data from real datasets, offering a powerful tool for data generation. However, it's important to note that GAN methodologies lack robust privacy guarantees [11]. Conversely, previous approaches have attempted to generate synthetic data using various methodologies. Some rely on summary statistics extracted from the original dataset, while others leverage specific domain-knowledge for data creation [12,13]. However, these conventional methods are bound by constraints, they operate within low-dimensional feature spaces, cater to specific fields, and notably lack the crucial aspect of differential privacy, making them less suited for contexts demanding robust privacy preservation measures. This integration of data synthesis with DP and the refinement of GAN techniques represents a substantial advancement, offering a promising pathway to create synthetic data that not only maintains utility but also ensures stringent privacy protections for individuals involved in the dataset.

The authors in [14] proposed a new strategy for privacypreserving data sharing called PrivateSMOTe. They used PrivateSMOTe in the highest-risk cases regarding k-anonymity. Differential privacy, a method for preserving privacy, is widely applied in various computer science domains. It's notably used in recommendation systems to maintain the confidentiality of user preferences and actions, enabling precise suggestions [15,16]. In data mining, differential privacy allows analyzing sensitive information without disclosing individual data [17]. Its application extends to crowd-sourcing [18], network measurements for protecting individual network traffic data while obtaining aggregate measurements [19], and in areas like intelligent transportation and sensor network stream processing [20]. The authors in [21] presents a thorough examination of differential privacy (DP) applications within smart city frameworks, particularly focusing on edge computing, to address the underexplored area of data privacy preservation in IoT-driven, resource-limited environments. This study [22] introduces an efficient, locally differentially private data aggregation scheme for smart grids, enhancing user privacy without incurring significant computational overhead, thereby addressing the limitations of current homomorphic encryption and randomization techniques. This paper [23] conducts a comprehensive literature review on the application of differential privacy to building data, exploring the current state, challenges, and future research opportunities in utilizing this privacy-preserving technique to enhance data utility while addressing privacy concerns.

However, these methods often rely on a trusted third party to gather data, apply algorithms, and conduct privacy-preserving analysis by adding "noise," leading to reduced data accuracy. Our approach enhances the synthesis framework, integrating stringent privacy assurances and rendering the synthetic data differentially private relative to the original dataset.

We implement differential privacy at the data source for increased privacy, complemented by the SMOTe algorithm and Gaussian noise addition for data generation. In studies such as [24–28], various researchers have proposed methodologies that utilize data masking techniques. Adding extra data masking in privacy-preserving methodologies can lead to several disadvantages:Reduced Data Utility, Increased Complexit and can slow down data processing, affecting the overall system performance. This is further strengthened by data masking techniques, ensuring data remains inaccessible to unauthorized entities without specific masking knowledge. These strategies, combined with differential privacy, offer robust privacy protection, allowing valuable computations on sensitive data while safeguarding individual privacy. Controlled noise addition hinders specific individual identification in datasets, yet permits meaningful data analysis and insights.

## 3. Preliminaries

This section provides background information on Synthetic Minority Over-sampling Technique (SMOTe) and the Gaussian noise approach. It also discusses K-anonymity which is a privacy concept and data anonymization method used to protect the identity and sensitive information of individuals in datasets. In addition, it investigates the implemented machine learning methods, such as K-nearest neighbours (KNN), Support vector machines (SVMs), and Naive bayes (NB). In the final section of this discussion, we will investigate the performance and evaluation metrics used to evaluate the performance of the proposed scheme.

### 3.1. Synthetic Minority Over-sampling technique (SMOTe)

SMOTe is a widely adopted method in the realms of machine learning and data mining, specifically designed to rectify class imbalance issues in datasets. Beyond its utility in class rebalancing, SMOTe finds applications in privacy-preserving data sharing through synthetic data generation. In the context of class imbalance, it effectively augments underrepresented class samples by creating synthetic instances based on existing data points. Simultaneously, in privacy settings, SMOTe contributes to safeguarding sensitive information by generating synthetic data that mirrors the statistical characteristics of the original dataset, enhancing privacy while still facilitating meaningful analysis and sharing of data across domains. Here are how it works:

- Identify minority class instances: SMOTe starts by identifying instances from the minority class that are candidates for generating synthetic examples. These instances serve as the basis for creating synthetic samples.
- Select nearest neighbors: For each minority class instance selected, SMOTe identifies its k-nearest neighbors among the minority class instances. The value of 'k' is a user-defined parameter that determines the number of nearest neighbors to consider.
- Generate synthetic samples: SMOTe generates synthetic examples by randomly selecting one of the k-nearest neighbors and interpolating between the feature values of the selected instance and the current instance. The interpolation is controlled by a random value between 0 and 1. For example, if k = 5 and the random value is 0.3, SMOTe creates a new instance by taking 30% of the feature values from the selected neighbor and 70% from the current instance.
- Repeat for all minority class instances: Steps 2 and 3 are repeated for all minority class instances selected in first step. This process generates multiple synthetic instances for each original minority class instance.

- Combine original and synthetic data: The synthetic instances are combined with the original data to create a balanced dataset. This balanced dataset can then be used for training machine learning models.

### 3.2. Gaussian noise

In the context of privacy-preserving techniques, "gaussian noise" typically refers to a method used to add noise to data in order to protect the privacy of individuals while still allowing for meaningful analysis. This technique is often employed in the field of differential privacy, which is a framework for preserving privacy in statistical and data analysis [29].

Consider $X_1, X_2, \ldots, X_N$ as a set of $N$ Gaussian random variables. These variables can be collectively referred to as a multivariate Gaussian distribution. The joint probability density function of these variables is given by:

$$F_X(\mathbf{x}) = \frac{1}{\sqrt{(2\pi)^N \det(\mathbf{K}_X)}} \exp\left(-\frac{1}{2}(\mathbf{x} - \boldsymbol{\mu}_X)^T \mathbf{K}_X^{-1}(\mathbf{x} - \boldsymbol{\mu}_X)\right), \quad (1)$$

where $\boldsymbol{\mu}_X$ is the mean vector and $\mathbf{K}_X$ is the covariance matrix of the distribution, defined as:

$$\boldsymbol{\mu}_X = \begin{bmatrix} \mu_1 \\ \mu_2 \\ \vdots \\ \mu_N \end{bmatrix}, \quad \mathbf{K}_X = \begin{bmatrix} K_{11} & \cdots & K_{1N} \\ K_{21} & \cdots & K_{2N} \\ \vdots & \ddots & \vdots \\ K_{N1} & \cdots & K_{NN} \end{bmatrix}. \quad (2)$$

#### 3.2.1. Additive perturbation

The additive perturbation method generates perturbed data $\mathbf{Y}$ by adding random noise $\mathbf{Z}$ to the original data $\mathbf{X}$, expressed as:

$$\mathbf{Y} = \mathbf{X} + \mathbf{Z}. \quad (3)$$

The covariance matrix of the original dataset $\mathbf{X}$ is given by:

$$\mathbf{K}_X = \left[(\mathbf{X} - \boldsymbol{\mu})(\mathbf{X} - \boldsymbol{\mu})^T\right]. \quad (4)$$

The noise $\mathbf{Z}$ is a jointly Gaussian vector with a zero mean. Its covariance matrix is defined as:

$$\mathbf{K}_Z = \left[\mathbf{Z}\mathbf{Z}^T\right]. \quad (5)$$

## 3.3. K-anonymity

K-anonymity is a methodology in data privacy aimed at anonymizing datasets to shield individual identities and sensitive details. It assures that each person's data, when disseminated or shared, remains unidentifiable and merged with a minimum of $M$ others, reducing the possibility of pinpointing individuals in the dataset. K-anonymity is notably used in securing demographic or health data [30].

**Definition 1:** (Quasi-identifier) Consider a dataset $D(B_1, …, B_m)$. A quasi-identifier for $D$ is a subset of attributes $\{B, …, B_j\} \subseteq \{B_1, …, B_m\}$ that requires regulated release.

The aim is to enable data dissemination from $D$ while ensuring individual anonymity. This requires the information released to non-specifically correspond to at least $M$ individuals, where $M$ is determined by the data custodian, as per the subsequent specification.

**Definition 2:** (M-anonymity Requirement) Each data release must ensure that for every quasi-identifier value combination, it matches non-specifically to at least $M$ individuals.

**Definition 3:** (M-anonymity) Let $D(B_1, …, B_m)$ be a dataset and $QI_D$ its quasi-identifiers. $D$ adheres to M-anonymity if, for every quasi-identifier $QI \in QI_D$, each value sequence in $D[QI]$ appears with at least $M$ instances in $D[QI]$.

Here, $QI_D$ signifies the set of quasi-identifiers linked to $D$, and $D$ indicates the cardinality, i.e., the count of tuples in $D$.

## 3.4. Machine learning techniques

**The K-nearest neighbors (KNN) classifier** is a fundamental yet vital algorithm in Machine Learning, used extensively in pattern recognition, data mining, and intrusion detection [31]. KNN identifies the nearest elements or clusters for a given query element, requiring a distance metric, given by:

$$d(z,v) = \left(\sum_{j=1}^{m} (z_j - v_j)^q\right)^{\frac{1}{q}} \quad (6)$$

**Support vector machines (SVM)** are robust supervised learning algorithms for classification. They find an optimal hyperplane in a multidimensional space to distinguish data classes, aiming to maximize the margin, the distance from the hyperplane to the nearest data points of each class [32].

Hyperplane Equation in SVM:

$$u \cdot z + c = 0 \quad (7)$$

where: $u$ is a weight vector orthogonal to the hyperplane. $z$ is a feature vector of a data point. $c$ is the hyperplane's offset from the origin along $u$.

Classification in SVM:

$$g(z) = sign(u \cdot z + c) \quad (8)$$

where $sign(.)$ is the sign function, returning -1 or 1 based on its argument's sign.

**Naive Bayes Classifier (NB):** These algorithms apply Bayes' theorem under the assumption of conditional independence between feature pairs, given a class variable. For class variable $y$ and dependent features $z_1$ through $z_m$:

$$P(y|z_1,…,z_m) = \frac{P(z_1,…,z_m|y)P(y)}{P(z_1,…,z_m)} \quad (9)$$

Assuming conditional independence:

$$P(z_j|y,z_1,…,z_{j-1},z_{j+1},…,z_m) = P(z_j|y) \quad (10)$$

Simplified relationship:

$$P(y|z_1,…,z_m) = \frac{P(y)\prod_{j=1}^{m} P(z_j|y)}{P(z_1,…,z_m)} \quad (11)$$

Classification decision in NB:

$$\widehat{y} = \underset{y}{argmax} \frac{P(y)\prod_{j=1}^{m} P(z_j|y)}{P(z_1,…,z_m)} \quad (12)$$

## 3.5. Performance evaluation measurements

The evaluation of the classifier's performance was conducted using Precision, Recall, and F-measure, derived from the confusion matrix detailed in Table 1. An explanation of these metrics is provided below.

- **Precision:** This metric evaluates the proportion of correctly identified actions among all selected actions. It's calculated as the ratio of correct selections to the total selections made.

$$Precision = \frac{TP}{(TP + FP)} \quad (13)$$

- **Recall:** This measures the fraction of accurately identified actions out of all the relevant actions.

Table 1. Confusion matrix.

|             | P′ (Predicted) | N′ (Predicted) |
|-------------|----------------|----------------|
| P (Actual)  | TP             | FN             |
| N (Actual)  | FP             | TN             |

Essentially, it assesses the percentage of relevant actions that have been correctly captured.

$$Recall = \frac{TP}{(TP+FN)} \qquad (14)$$

- **F-measure:** This is a composite metric that combines precision and recall, providing a weighted average of the two. It's particularly useful for assessing the balance between precision and recall.

$$F-measure = \frac{2*Precision*Recall}{(Precision+Recall)} \qquad (15)$$

## 4. The proposed privacy-preserving model in smart homes

This section describes the proposed approach and model, including a thorough explanation of the data capture procedure, applied preprocessing methodologies, and any specific algorithms or techniques that are intricately incorporated into the model's framework as shown in Fig. 1.

### 4.1. System model and overview

Based on the basic idea of differential privacy and SMOTe, the proposed approach consists of two steps: data synthesis and noise addition. This approach proposes a differential privacy-based system to protect the privacy of data collected from smart homes by adding synthetic data followed by noise using SMOTe algorithm. Given an initial data $D_o$ for smarthome owner, the objective is to protect the privacy of data by generating a synthetic dataset $D_s$ that closely matches the statistical characteristics of $D_o$.

### 4.2. Problem formulation

Consider the dataset $D_o$ which consists of $n$ records, each containing $d$ characteristics. A synthetic dataset $D_s$ is considered when it closely resembles $D_o$ across all functions, as indicated by $f(D_s) = f(D_o)$. Our investigation focuses on statistical metrics classification models.

### 4.3. Generating synthetic data and apply Gaussian noise

In this research phase, we start by thoroughly understanding the smart home dataset, focusing on attributes with sensitive information, which are anonymized and generalized for privacy. The widely used SMOTe technique, originating from machine learning and data mining, addresses class imbalance and privacy concerns. It generates synthetic data that maintains the statistical features of the original dataset, enhancing privacy and enabling meaningful cross-domain data analysis and sharing. as described in section 3. The algorithm delineated in Algorithm 1 articulates the methodology underlying the proposed approach. This approach encompasses the following steps:

- Identification of Minority Class Instances: Initial identification of minority class instances to generate synthetic samples.
- Selection of Nearest Neighbors: Each minority class instance has its k-nearest neighbors identified within the same class, with $k$ being a predefined parameter.

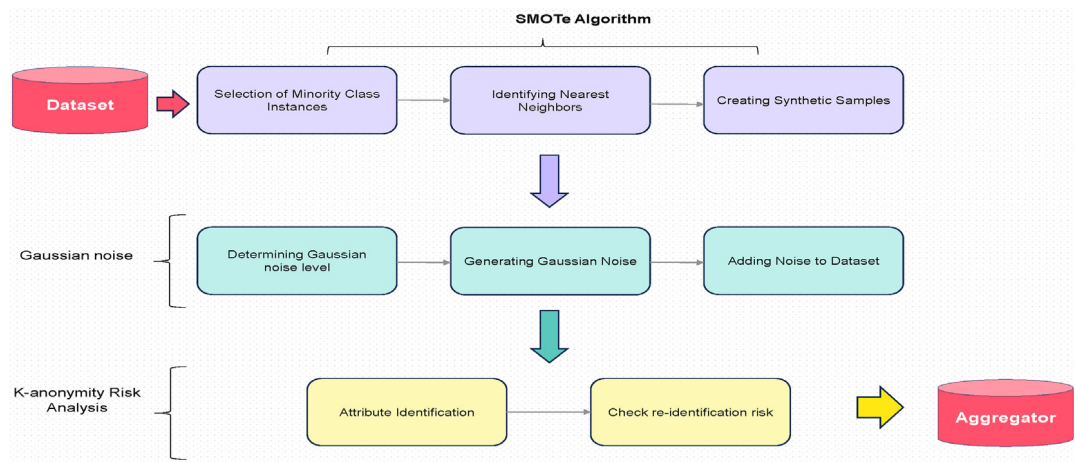

Fig. 1. The proposed system model.

- Synthetic Sample Generation: Synthetic samples are produced through interpolation between the feature values of a minority class instance and one of its k-nearest neighbors, chosen at random.
- Repetition Across All Instances: This neighbor selection and sample generation process is repeated for each instance in the minority class.
- Data Integration: The resultant synthetic instances are integrated with the original dataset, creating a balanced dataset for model training.
- Additive Perturbation Phase: The dataset undergoes Gaussian noise addition, a differential privacy technique enhancing individual privacy while allowing valuable data analysis.

**Computational Complexity Analysis:** Lines 7 to 9 finding the s nearest neighbors for each minority sample can be time-consuming. If a simple linear search is used, this step could take $O(M*N*d)$ time, where M is the number of minority samples, N is the total number of samples, and d is the number of dimensions/attributes. Lines 10 to 15: the time complexity for generating each synthetic sample is $O(d)$, since it involves a calculation for each attribute. The outer loop runs E times, making this step $O(E*d)$. Line 17: merging datasets is generally a linear operation, $O(M + E)$, assuming E synthetic samples are created. Line 19: adding Gaussian noise involves generating a random value for each attribute of

---

**Algorithm 1:** Enhanced SMOTe(M, E, s)

**Parameter:** $D$ Original data; Count of minority samples $M$; Amount of SMOTe $E\%$;
   Neighbor count $s$, $K_j$ Noise for generating $V_j$.
**Input:** $Original[\ ][\ ]$: Array for original minority samples
   $synthIndex$: Counter for synthetic samples, initialized to 0
   $Enhanced[\ ][\ ]$: Array for synthetic samples.
**Output:** $V_j$: Adjusted copy of $D$ with trust level $j$ combined with $(E/100) * M$ enhanced minority samples

/* If E is below 100%, shuffle minority samples as only a portion will undergo Enhanced SMOTe. */
1  **if** $E < 100$ **then**
2  |  Shuffle $M$ minority samples $M = (E/100) * M$
3  |  $E = 100$
4  $E = (E/100)$ /* Assume Enhanced SMOTe amount in multiples of 100. */
5  $s$ = Neighbor count
6  $attributeCount$ = Number of attributes
   /* Compute s nearest neighbors for each minority sample. */
7  **for** $j = 1$ to $M$ **do**
8  |  Compute $s$ nearest neighbors for $j$, store indices in $neighborArray$
   |  GenerateSynth($E, j, neighborArray$)
   /* Function to create synthetic samples. */
9  $GenerateSynth(E, j, neighborArray)$ **while** $E \neq 0$ **do**
10 |  Select a random number between 1 and $s$, name it $nn$.
11 |  **for** $attr = 1$ to $attributeCount$ **do**
12 |  |  Calculate: $difference = Original[neighborArray[nn]][attr] - Original[j][attr]$
13 |  |  Calculate: $randomGap$ = random number between 0 and 1
14 |  |  $Enhanced[synthIndex][attr] = Original[j][attr] + randomGap * difference$
15 |  $synthIndex++$
16 |  $E = E - 1$
17 Merge $D$ with $(E/100) * M$ synthetic samples
18 Create $W$ where $W = K_j D$ /* Add Gaussian noise. */
19 Generate $V = D + W$
20 Return $V$: /* End of GenerateSynth. */

each synthetic sample, which has a complexity of $O(E*d)$.

Therefore, the overall time complexity of the algorithm is dominated by the nearest neighbors' calculation and the generation of synthetic samples. The exact time complexity would depend on the method used for nearest neighbor search and the size of the dataset, but assuming a linear search, the complexity would be approximately $O(d*(M*N+E))$.

## 5. Experimental and analysis

To evaluate the effectiveness of the proposed methodology, this study utilizes a substantial real-world dataset, namely the MHEALTH dataset [33]. The dataset consisting of approximately 1 million records. The data primarily consists of numerical values. Specifically, it is referred to as the "Mobile HEALTH" dataset, which captures body motion and vital signs recordings. The dataset encompasses measurements from ten volunteers with diverse profiles while engaging in various physical activities.

Additionally, the research employs Google Colab [34], a free, cloud-based platform that provides a Jupyter notebook environment for writing and executing Python code. It offers easy access to powerful computing resources like GPUs and TPUs, making it ideal for machine learning and data analysis projects. Colab facilitates seamless collaboration and integrates with Google Drive for convenient file storage and sharing. Table 2 presents the configurations for the *XGBoost* and *KNN* (with $K = 3$) algorithms. These algorithms are integral to the proposed method and are utilized for analyzing data shared in smart home environments.

### 5.1. Classification evaluation

This subsection analyses the influence of different parameters on the classification results, including the Gaussian noise value *g*, the amount of SMOTe *N* % and the re-identification risk for *k* value when apply K-anonymity function. Knowing that the accuracy for the dataset without applying the proposal were DT: 98.11%, KNN: 98.4%, NB: 89.91% and XGBoost: 96.01%.

Figure 2 delineates the interplay between the accuracy of several classification algorithms and the extent of Gaussian noise, following the application of a SMOTe percentage of $N = 130\%$. The Gaussian noise ranges from 0 to 1, within which the accuracy for various classification methods is observed as follows: Decision Tree (DT) between 85.60% and 98.05%, K-Nearest Neighbors (KNN) from 87.82% to 98.74%, Naive Bayes (NB) within 86.55%–89.45%, and XGBoost achieving 86.38%–95.86%. Comparing these results to the baseline accuracies obtained without our proposed methodology, the error margin appears dependent on the Gaussian noise level, remaining relatively stable at $g = 0.3$ and increasing thereafter.

Table 2. XGBOOST Configuration.

| Parameter | Value |
| --- | --- |
| Number of Estimators | 61 |
| Minimum Child Weight | 7 |
| Maximum Tree Depth | 6 |
| Gamma Value | 0.4 |
| Number of Rounds | 10 |

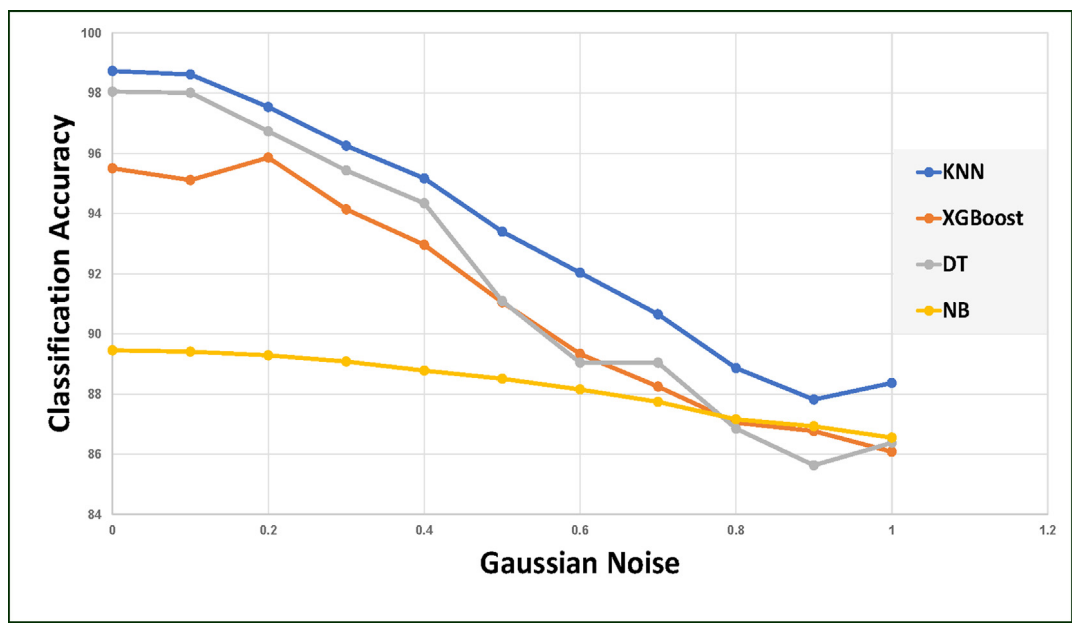

Fig. 2. Gaussian noise for amount of SMOTe $N = 130\%$ and $k = 2$.

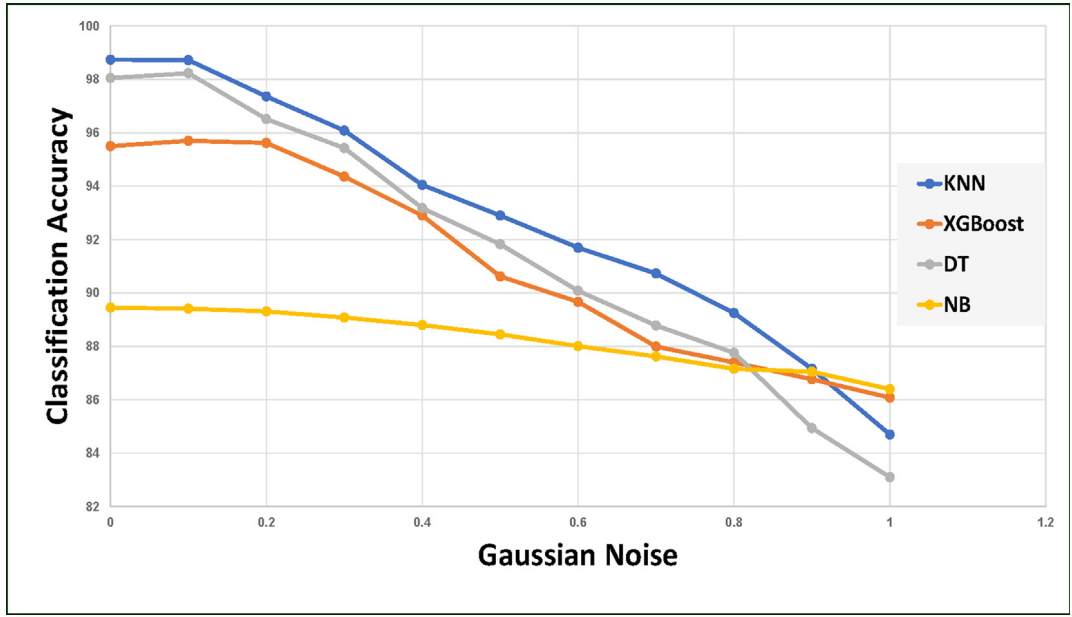

Fig. 3. Gaussian noise for amount of SMOTe N = 220% and k = 2.

Figure 3 illustrates the relationship between the accuracy of various classification methods and Gaussian noise levels after applying a SMOTe percentage of $N = 220\%$. With Gaussian noise fluctuating between 0 and 1, the resulting accuracies for the algorithms are as follows: DT with 83.01%–98.23%, KNN between 84.7% and 98.74%, NB ranging from 86.40% to 89.45%, and XGBoost achieving 86.08%–95.7%. These outcomes, when contrasted with the accuracies attained without our proposed approach, demonstrate an error margin linked to the level of Gaussian noise, particularly stable at $g = 0.3$ but escalating with higher noise levels.

In Fig. 4, the correlation between the accuracy of different classification algorithms and Gaussian noise, under a SMOTe percentage of $N = 370\%$, is presented. The Gaussian noise, varying from 0 to 1, impacts the accuracy of the algorithms as follows: DT records accuracy between 84.70% and 98.26%, KNN shows 86.73%–98.74%, NB operates within 86.34%–89.45%, and XGBoost achieves 85.82%–95.68%. When these results are compared with the accuracies obtained without the implementation of our methodology, the error margin, influenced by the Gaussian noise, remains relatively stable at $g = 0.3$ and increases with higher levels of noise.

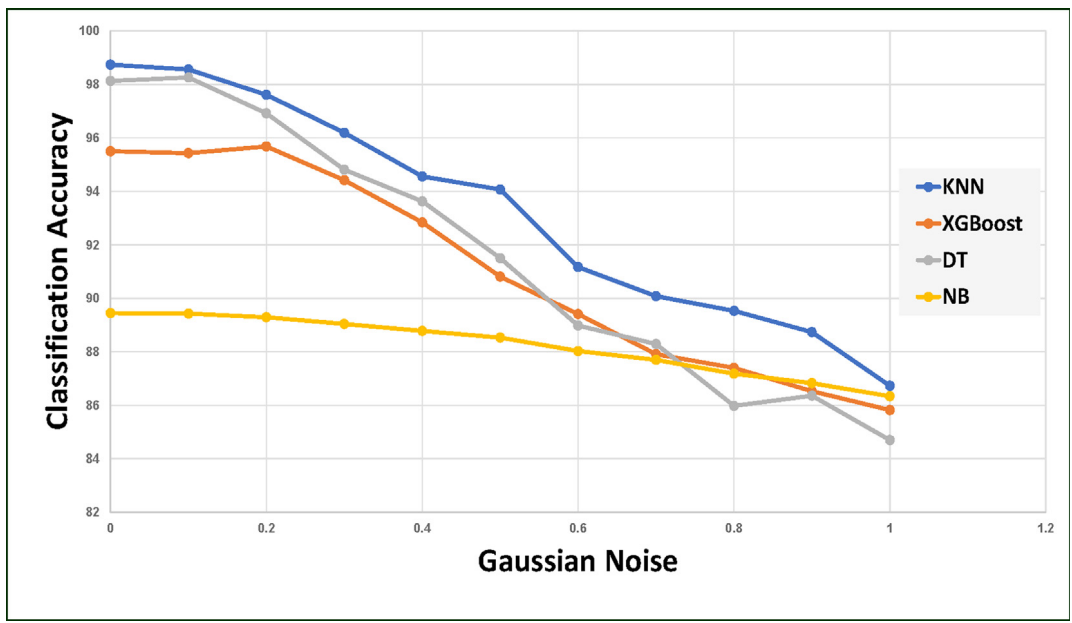

Fig. 4. Gaussian noise for amount of SMOTe N = 370% and k = 2.

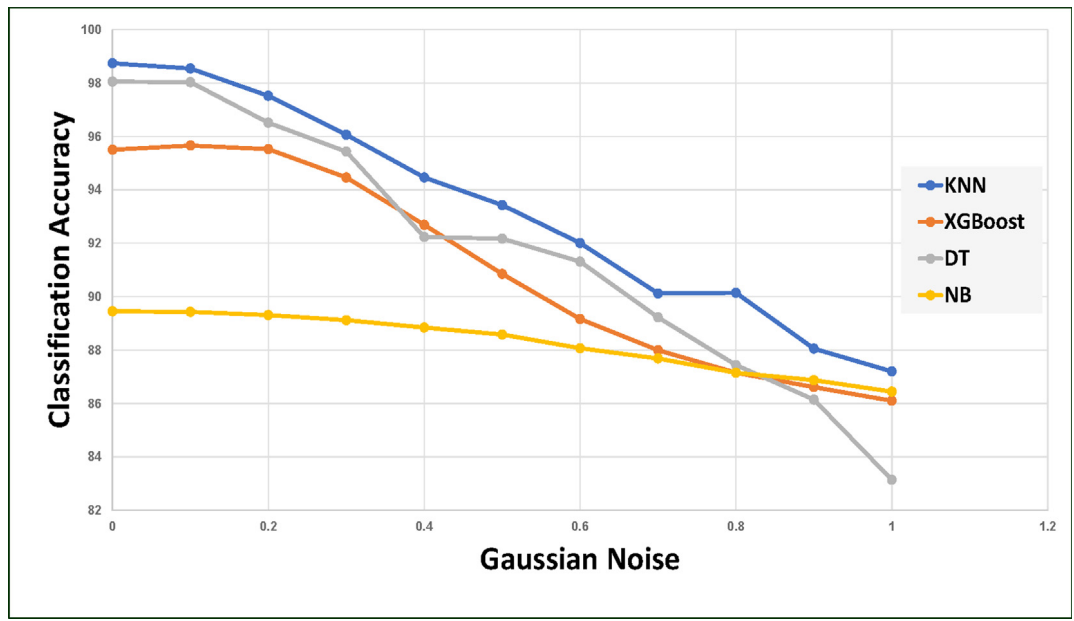

Fig. 5. Gaussian noise for amount of SMOTe $N = 500\%$ and $k = 2$.

Figure 5 expounds on the relationship between the accuracy of various classification algorithms and Gaussian noise after the application of a SMOTe percentage of $N = 500\%$. Within the Gaussian noise range of 0–1, the accuracies for the algorithms are observed as follows: DT maintains 83.14%–98.05%, KNN ranges between 87.2% and 98.74%, NB records 86.44%–89.45%, and XGBoost achieves 86.1%–95.66%. Comparing these findings with the baseline accuracies obtained without our methodology, the error margin is seen to depend on the Gaussian noise level, showing a relative stability at $g = 0.3$ before increasing at higher noise levels.

Our comprehensive analysis, as depicted in previous figures, reveals a consistent trend in the interplay between the accuracy of classification algorithms and the level of Gaussian noise, across varying SMOTe percentages $N = (130\%, 220\%, 370\%, and\ 500\%)$ Irrespective of the SMOTe percentage applied, the Gaussian noise, ranging from 0 to 1, uniformly affects the accuracy of the classification methods under study - Decision Tree (DT), K-Nearest Neighbors (KNN), Naive Bayes (NB), and XGBoost. These effects are evident within specific accuracy ranges for each algorithm. A pivotal observation is that the error margin, which is linked to the level of Gaussian noise, exhibits a similar pattern across all SMOTe levels. Particularly, this error margin remains relatively stable at a Gaussian noise level of $g = 0.3$, and then progressively increases with higher noise levels. This trend underscores the robustness and consistency of our proposed methodology in maintaining accuracy while ensuring privacy, despite variations in data augmentation via SMOTe.

5.2. Security analysis

This section delves into the utilization of the k-anonymity model as a pivotal framework for conducting security analysis. The k-anonymity approach is fundamentally designed to thwart the re-identification of individuals within a dataset. It accomplishes this by ensuring that within any grouping of k records, the shared attributes of the individuals are sufficiently homogenized. This level of similarity provides a cloak of anonymity, effectively safeguarding individual identities.

K-anonymity is both a statistical and computational strategy that plays a crucial role in the defense of sensitive data. Its primary objective is to create a formidable barrier against the efforts of potential adversaries or external entities who might attempt to isolate and extract specific personal details pertaining to individuals from a dataset. By blending individual data points into larger, indistinguishable groups, k-anonymity makes the task of pinpointing unique personal information significantly more challenging, thereby enhancing the overall security and confidentiality of the data. This methodology, as discussed in the seminal work of Samarati (2001), has emerged as a cornerstone in the field of data privacy and protection [35].

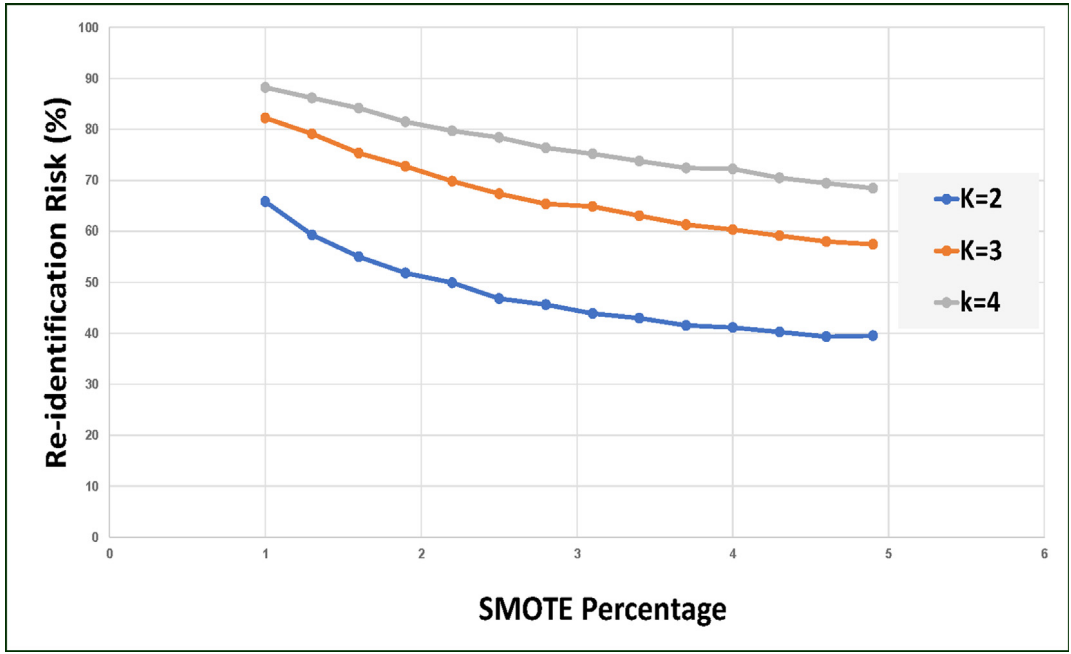

Fig. 6. Re-identification risk and SMOTe relationship when gaussian noise g = 0.1.

In Fig. 6, the interrelation between the re-identification risk and varying SMOTe amounts, set against a Gaussian noise level of $g = 0.1$, is expertly depicted. The data portrays a clear trend: with an increase in the SMOTe amount, there is a notable reduction in the re-identification risk. This consistent pattern at a Gaussian noise level of $g = 0.1$ underlines the efficacy of SMOTe in mitigating privacy risks.

Figure 7 effectively illustrates the relationship between re-identification risk and the application of different SMOTe amounts, within the context of a Gaussian noise level of $g = 0.3$. The observed trend remains consistent, indicating that as the SMOTe amount is augmented, the re-identification risk correspondingly diminishes. This trend at a Gaussian noise level of $g = 0.3$ reaffirms the role of SMOTe in enhancing data privacy.

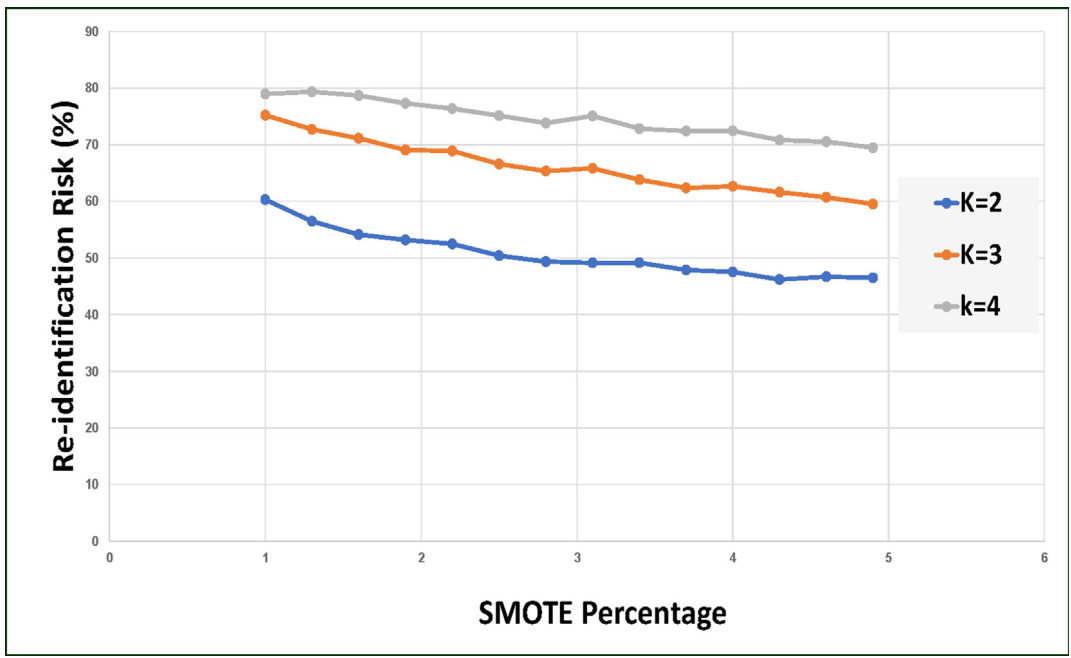

Fig. 7. Re-identification risk and SMOTe relationship when gaussian noise g = 0.3.

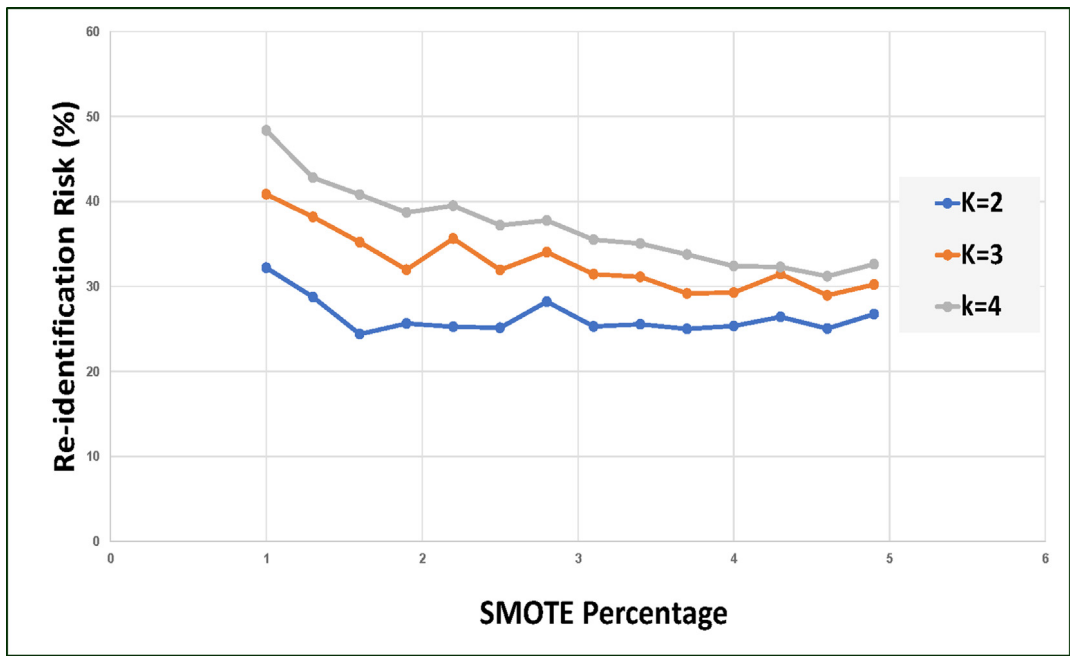

Fig. 8. Re-identification risk and SMOTe relationship when gaussian noise $g = 0.6$.

In the context of a Gaussian noise level of $g = 0.6$, Fig. 8 presents a detailed analysis of the correlation between the re-identification risk and various SMOTe amounts. The trend observed here aligns with previous findings, showcasing a decrease in the re-identification risk as the SMOTe amount increases. This pattern, consistent at a Gaussian noise

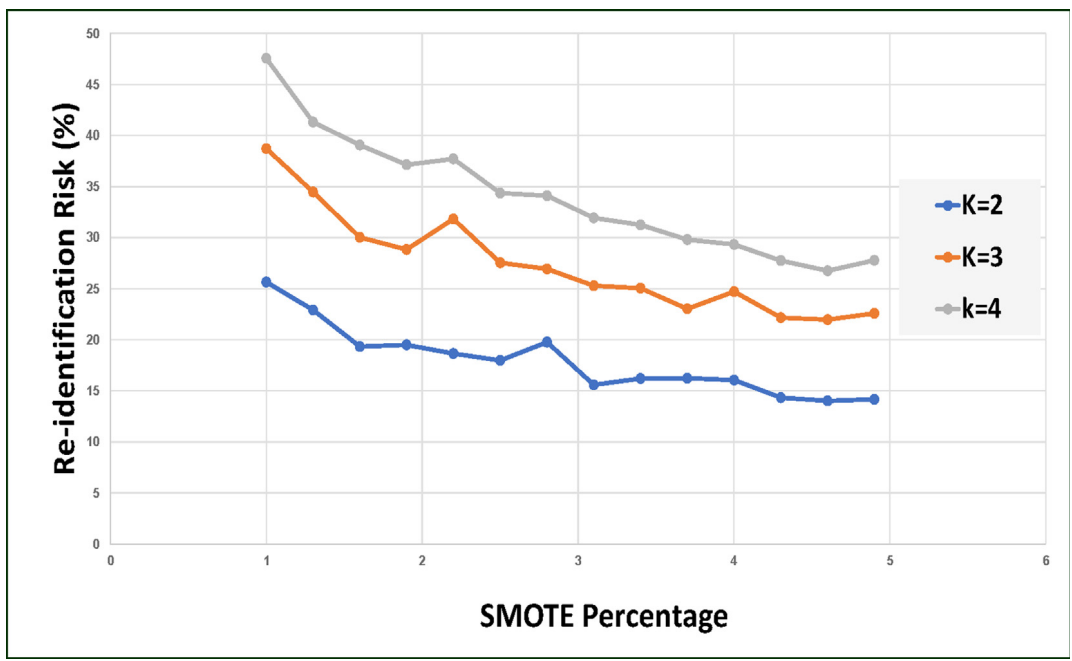

Fig. 9. Re-identification risk and SMOTe relationship when gaussian noise $g = 1.0$.

level of $g = 0.6$, underscores the effectiveness of SMOTe in reducing privacy vulnerabilities.

Figure 9 explores the relationship between the re-identification risk and different SMOTe amounts at a Gaussian noise level of $g = 1$. The figure reveals a trend consistent with lower levels of Gaussian noise: an increase in the SMOTe amount leads to a decrease in the re-identification risk. This trend, observed at the highest Gaussian noise level of $g = 1$, further validates the utility of SMOTe in the realm of privacy preservation. Each variation focuses on the consistent trend observed at different Gaussian noise levels, emphasizing the effectiveness of SMOTe in reducing re-identification risk across varying scenario.

The previous results indicate that the most effective outcomes were attained by setting Gaussian noise to 0.3 and increasing the SMOTe ratio to 500%, effectively balancing data accuracy and privacy. The implementation of a k-anonymity function successfully mitigated re-identification risks. This combination of parameters led to a marked enhancement in data security and user privacy within smart home settings.

## 6. Conclusion

This study presents a novel approach for secure data sharing in smart homes, prioritizing user privacy through an innovative method that combines data synthesis, SMOTe algorithm, and Gaussian noise application. Our findings demonstrate that this technique effectively safeguards user privacy, as evidenced by high privacy standards, accuracy, recall rates, and f-measure metrics. The strategy is particularly notable for achieving significant utility in privacy by maintaining high classification accuracy and markedly reducing re-identification risks. Specifically, optimal results were observed with a re-identification risk at 30%, Gaussian noise $g = .3$ and amount of SMOTe $N = 500\%$ and $k = 2$ applied alongside k-anonymity function. The classification accuracy ranged between 90% and 98% for the utilized classification techniques. This research advances data privacy and utility in smart homes and suggests a promising framework for privacy-preserving solutions in the broader Internet of Things (IoT) domain. Future work will explore extending these methods to multimodal data in smart home contexts, integrating information from diverse sensors and devices.

## Conflicts of interest

The authors declare no conflict of interest.